\documentclass[useAMS,usenatbib]{mn2e}

\usepackage{epsf}
\usepackage{epsfig}
\usepackage{amssymb}
\usepackage{amsfonts}

\title[Results from the AO coronagraph at the WHT]{Results from the adaptive optics coronagraph at the WHT}
\author[S. J. Thompson, A. P. Doel et al.]{S. J. Thompson$^{1}$\thanks{E-mail:
sjt@star.ucl.ac.uk (SJT)}, A. P. Doel$^{1}$, R. G.
Bingham$^{1}$, A. Charalambous$^{1}$, R. M. Myers$^{2}$,
\newauthor
N. Bissonauth$^{2}$, P. Clark$^{2}$, G. Talbot$^{3}$ \\
$^{1}$Department of Physics and Astronomy, University College London, Gower Street, London, UK\\
$^{2}$Centre for Advanced Instrumentation, Durham University, Dept of Physics, South Road, Durham, UK\\
$^{3}$Isaac Newton Group of Telescopes, Apartado de correos 312, E-38700 Santa Cruz de la Palma, Tenerife, Spain}
\begin{document}

\date{Accepted 2005 September 14. Received 2005 September 1; in original form 2005 June 17}

\pagerange{\pageref{firstpage}--\pageref{lastpage}} \pubyear{2005}

\maketitle

\label{firstpage}

\begin{abstract}
Described here is the design and commissioning of a coronagraph facility for
the 4.2 metre William Herschel Telescope (WHT) and its Nasmyth Adaptive Optics system
for Multi-purpose Instrumentation (NAOMI).  The use of the NAOMI system gives an
improved image resolution of $\sim 0.15$ arcsecs at a wavelength of 2.2$\mu$m.
This enables the \underline{O}ptimised \underline{S}tellar \underline{C}oronagraph
for \underline{A}daptive optics (OSCA) to suppress stellar light using smaller occulting
masks and thus allows regions closer to bright astronomical objects to be imaged.  OSCA
provides a selection of 10 different occulting masks with sizes of 0.25 - 2.0
arcsecs in diameter, including two with full greyscale Gaussian profiles.
There is also a choice of different sized and shaped Lyot stops (pupil plane masks).
Computer simulations of the different coronagraphic options with the NAOMI segmented mirror
have relevance for the next generation of highly segmented extremely large telescopes.
\end{abstract}


\section{Introduction}

Compared to a standard coronagraphic imaging system, one with adaptive optics
(AO) gives much higher spatial resolution and a high dynamic range allowing the
environments of bright objects to be studied closer in than ever before \citep{mal96}.
Additional instrumentation used in combination with a coronagraph allows
other new areas of research to be pursued.  There are few
coronagraphs that have this facility, e.g. CIAO on Subaru, which has a choice of
linear polarimeters \citep{mur03} and OSCA on the WHT, which has a spectroscopic capability
(with the OASIS integral field unit).

The adaptive optics system at the WHT, NAOMI, consists of a single Shack-Hartmann wavefront
sensor normally using $8\times8$ sub-apertures and a 76-element segmented mirror.
This is a reasonably high-order AO system and can offer partial correction for wavelengths
down to 700nm.  NAOMI has been at the telescope since 2000 and during early 2003 was
moved to a new environment-controlled laboratory at the opposite Nasmyth platform.  This
provides a more dust-free and thermally stable environment and so should
improve the performance of the system \citep{mye03}.

OSCA is a fully deployable instrument which when in use leaves the focus of the
NAOMI beam unchanged.  This enables OSCA to be used in conjunction with a
number of instruments that have already been commissioned at the WHT (Fig. 1).  The main
imaging camera used with OSCA is the Isaac Newton Group Red Imaging Device (INGRID);
a $1024 \times 1024$ element HgCdTe
cooled near-IR detector at the NAOMI focus.  The pixel scale when used with
NAOMI is $\sim$0.04 arcsec/pixel, hence Nyquist sampling is only obtained down to
H-band ($1.6 \mu$m).  Prior to the detector but within the camera cryo-chamber
are a set of 3 wheels containing broadband filters ({\it Z - K}),
narrowband filters and a choice of pupil stops respectively.  OSCA also has the
option of being used in conjunction with an integral field spectrograph (OASIS) for
imaging and spectral analysis at visible wavelengths.

\begin{figure*}
\begin{center}
\epsfig{file=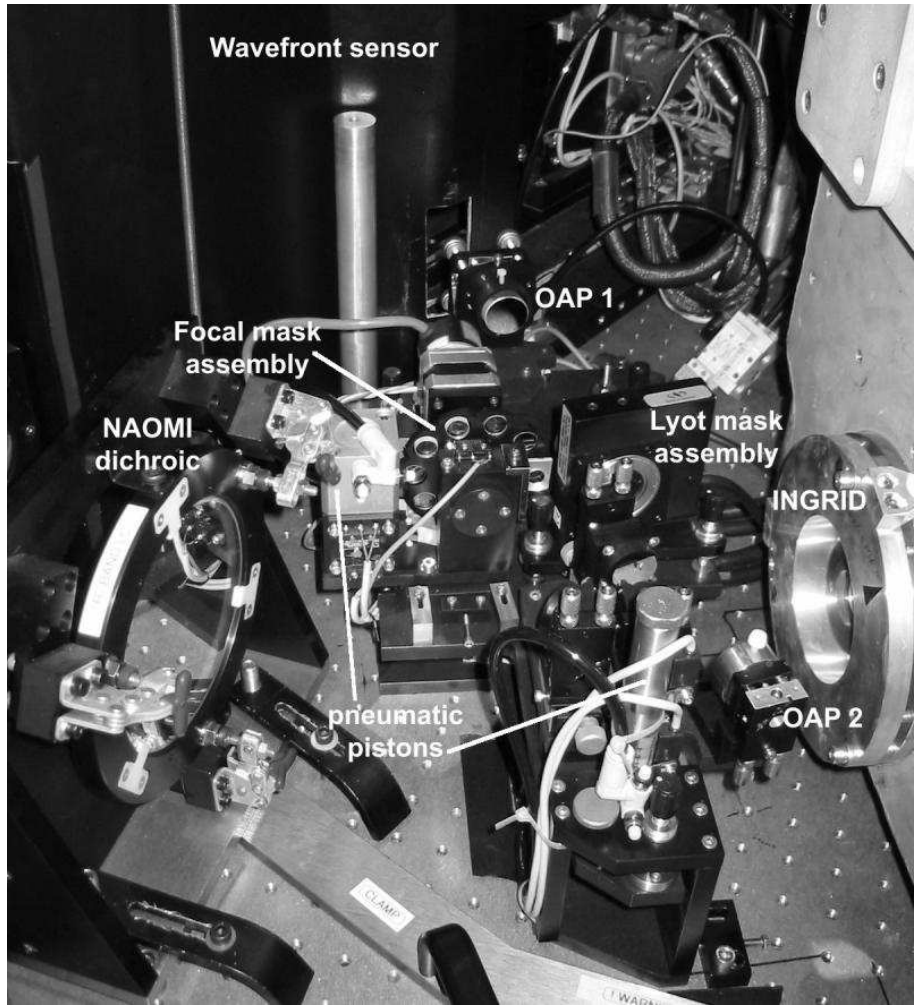,width=12cm,bbllx=20pt,bblly=20pt,bburx=308pt,bbury=339pt}
\caption{OSCA installed at the GRound based Adaptive optics Controlled Environment
(GRACE) at the William Herschel Telescope.  The light
from NAOMI enters from the left through the dichroic and exits to the INGRID
detector shown at the right-hand side of the picture.  For scale, the hole spacing on the
optical table is 25mm.}
\label{osca_grace}
\end{center}
\end{figure*}
An important criterion in creating a high contrast imaging system is
keeping scattered light to a minimum.  Compared to other AO
coronagraphs, the system at the WHT allows the insertion of an
occulting mask in a focal plane before the AO system as well as those in the
focal plane within the AO.  Ideally this `pre-AO' mask would be made of a
dichroic material that is transparent to the wavelengths used for wavefront
sensing and opaque to the science/observation wavelength.
The many square segments in the NAOMI adaptive mirror contribute more scattered
light than a similar sized continuous face-sheet mirror and the gaps between the
segments also have a higher emissivity in the infra-red, so this
additional pre-AO stop for the NAOMI-OSCA system could be extremely useful
in cutting down scattered light at the science wavelengths and thus
improving sensitivity in the final image.  An additional means of reducing the
diffraction effects of the segmented mirror subsequent to the application of a
focal occulting mask is the use of a Lyot (pupil) mask that is matched to the
segmented mirror.  Such a mask is shown in Fig. \ref{lyot} (inset top, left-hand mask) and
discussed in \S 3.2.

OSCA was commissioned at the WHT in mid-2002 and became available for general
use in 2003b.  The OSCA + OASIS option is currently unavailable although work is
being planned to rectify the problem.

\section{Design overview}

OSCA \citep{tho03} is based on the classic Lyot \citep{ly39} coronagraph design.  The sizes of
the focal plane masks (0.25 - 2.00 arcsec) were chosen to take advantage of the improved
point-spread function (PSF) offered
by NAOMI.  Ten focal plane masks are available for selection on the instrument.
One is for alignment purposes only and two are of a more novel Gaussian profile
compared to the standard hard edged circular discs.

The standard focal plane (occulting) masks are made of chromium, deposited
on the substrates by a contact-photolithography process.
Computer simulations, both here (in \S \ref{simulations}) and by others \citep{nak94},
have shown that occulting masks of a Gaussian profile give better
suppression than the standard solid disc design.  However creating such a mask
is not trivial.  Canyon Materials Ltd have a patented glass formula, HEBS (High
Energy Beam  Sensitive)-glass, which behaves in a similar way to photographic
film.  HEBS-glass is sensitive towards electron-beam exposure whereby exposure with a
certain electron-beam dosage changes the optical density of the material.  In
this way the HEBS-glass can act as a mask material and by varying the electron-beam
exposure different greyscale (optical density) levels can be written into the
glass \citep{wal96}.  To create the Gaussian profile masks for OSCA
(Fig. \ref{gaus_masks})
669 different greyscale levels were written in ring steps of $0.5\mu$m width.
One restriction on their usage is that the glass is only suitable for the
wavelength range 0.4 - 0.8$\mu$m.  There is a visible camera on OASIS, so these
masks were included in OSCA with this use in mind.
\begin{figure}
\begin{center}
\epsfig{file=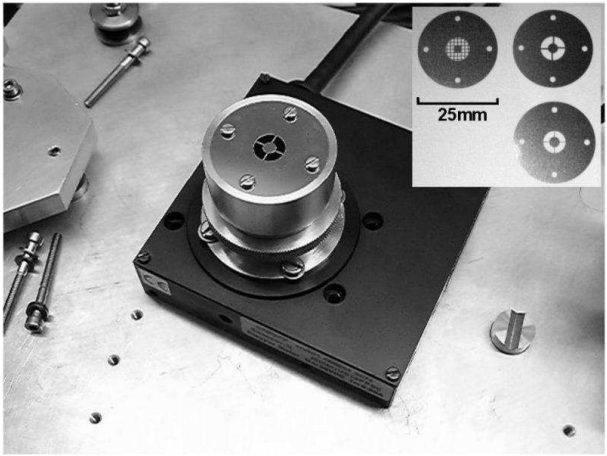,width=8cm,bbllx=20pt,bblly=20pt,bburx=308pt,bbury=236pt}
\caption{Standard OSCA Lyot mask mounted on a step-motor rotation stage.  Other
available Lyot masks are shown inset.}
\label{lyot}
\end{center}
\end{figure}

The pupil plane masks are manufactured from 0.25mm thick, hard stainless steel
by a process of photo-chemical machining.  Since OSCA is located at a Nasmyth
focus of the telescope, the telescope field derotator (which ensures the
observation object does not rotate as the telescope tracks on its alt-az
mounting) has the effect of rotating the telescope pupil.  Since the standard Lyot
mask used in OSCA includes vane masking, the Lyot mask must be rotated to
maintain the alignment with the telescope pupil image.  The standard Lyot mask
mounted on a Newport SR50 step motor rotation stage is shown in Fig. \ref{lyot}.
Additional pupil stops (non-rotating) were manufactured at the ING to go in the
INGRID pupil filter wheel.  These provide cooled masks for OSCA which are
particularly important for observing in {\it K}-band otherwise the thermal background
noise is very high.
\begin{figure}
\begin{center}
\epsfig{file=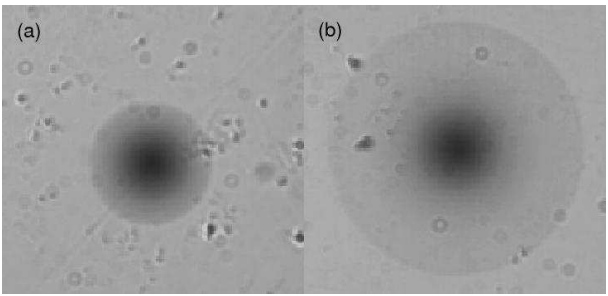,width=8cm,bbllx=20pt,bblly=20pt,bburx=283pt,bbury=146pt}
\caption{White light images of the OSCA Gaussian profiled HEBS occulting masks.
Speckles are dust in the CCD camera.}
\label{gaus_masks}
\end{center}
\end{figure}

An important requirement of the optical design was to ensure that the
focal position was unaltered with OSCA deployed in the beam path and since
space is very limited on the optical table,
to fit this design into a very small spatial envelope (see Fig. \ref{osca_grace}).
The optical specifications were that it should be diffraction limited at a wavelength
of 2.2$\mu$m over the full field of view and have less than 0.1 arcsec distortion
over the field for the full wavelength range.  Additionally, the system has an
exit pupil in the same position as that for NAOMI so providing an unchanged
optical interface to subsequent optics which were designed to work with NAOMI.
The field of view for OSCA is designed to be 20 arcsec,
but ghosting within the INGRID camera optics has currently reduced this to 15 arcsec.
The system also has a 1:1 magnification to leave the NAOMI field-scale unchanged.
To accommodate the required wavelength range of 0.4 - 2.4$\mu$m all mirrors
in the system were coated with protected silver on Zerodur.  Similarly
all transmissive optics were made from an infra-red (IR) grade fused silica
(Spectrosil WF) which gives over 90\% transmission across the entire
specified wavelength range and can be polished to a very high surface
quality.

\begin{figure}
\begin{center}
\epsfig{file=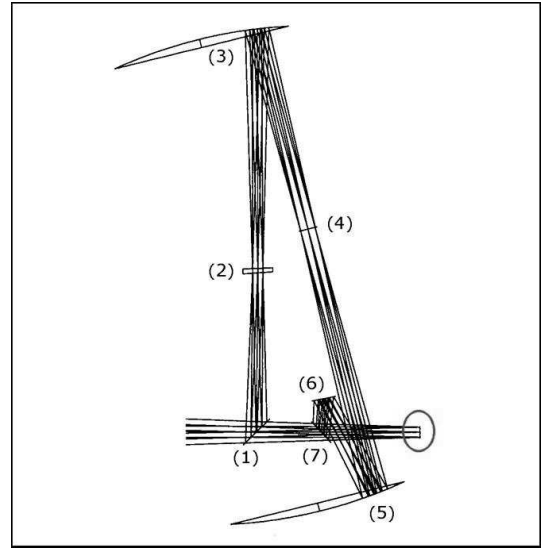,width=7cm,bbllx=20pt,bblly=20pt,bburx=180pt,bbury=186pt}
\caption{OSCA ray-trace.  The sequence of optical components is numbered and the
original NAOMI beam (without OSCA) is overlaid to show the focus at INGRID (circled)
unchanged.}
\label{zemax}
\end{center}
\end{figure}

To prevent ghosting within OSCA the substrates for the focal plane masks have
a slight wedge angle.  The wedges are also tilted slightly in the design,
to ensure the ghost beam misses the pupil.  This is achieved by aligning the
whole occulting mask filter-wheel assembly at a slight angle to the beam and
checking that the ghost-beam is thrown completely out of the optical path.

The path of light through OSCA is shown in Fig. \ref{zemax} and is numbered
sequentially along the beam direction.  When deployed in the NAOMI beam, flat
mirror (1) picks up the beam.  The occulting masks are located at the focal
plane (2).  The Lyot stop is located at the pupil plane (4).  The mirrors at (3)
and (5) are a pair of off-axis paraboloids, and finally the flats at (6) and (7)
position the focus to the original NAOMI focal position.

The important mechanical requirements of the OSCA system were for it to be
thermally stable from -5$^\circ$C to 25$^\circ$C (i.e. positional accuracy is held
within the specified tolerances and that the components will function correctly),
 vibrationally stable, for none of the components to infringe upon the beam path
within or outside of OSCA, the entire system to be deployable in and out of the
beam with a repeatability of $\pm50\mu$m, the centre of the focal plane masks to
be automatically positioned to $\pm10\mu$m, 5 year component lifetime (standard
ING requirement) and for the heat output to be minimised so as not to interfere with INGRID.

The unit can be automatically deployed to interrupt the beam passing
through to INGRID and the OASIS pick-off mirror.  A vertical deployment was
implemented due to the space restrictions.  From Fig. \ref{osca_grace} it can be seen that
OSCA comprises of a base plate which is rigidly clamped to the bench and a
top plate on which all the opto-mechanical components are seated.  Deployment is
achieved by the top plate pivoting about a
groove and cone of hardened stainless steel at the first off-axis paraboloid end of OSCA
with a pneumatic actuator at the second off-axis paraboloid end to bring the plate up
into the beam.

The focal masks are located within a wheel that can be automatically selected
from the control room.  Each focal mask is re-located to high precision by use
of notches around the mounting wheel and a pneumatically deployed arm which
locks into the notches (detent-arm).  This facility along with the capability to
deploy OSCA into and out of the beam accurately and automatically allows for a
flexible and versatile observing program over the course of a night.

\section{System simulations}
\label{simulations}

Computer simulations were written and run for a coronagraphic system with a
segmented adaptive optics corrector (as in NAOMI).  A program that generates
Kolmolgorov type turbulence phase-screens \citep{lan92} was used as the input to
the simulation.  This program does not incorporate the outer scale of atmospheric turbulence,
although this effect is negligible over a 4m aperture.

Unless stated otherwise, the simulations discussed here used a phase-screen
(aperture) size of $256 \times 256$ padded with zeros to four times its size for the fast
Fourier transforms (FFT), giving a $1024 \times 1024$ focal plane.  In the final image
this results in a higher resolution than that achieved by INGRID. The AO and coronagraph
code were developed in M{\sc atlab} \citep{tho04} based on an original C-code by A.P. Doel.
The simulations include photon noise in the wavefront sensor, mirror hysteresis and pixel
noise in the detector.

\subsection{Effect of focal mask properties on performance}

\begin{figure}
\begin{center}
\epsfig{file=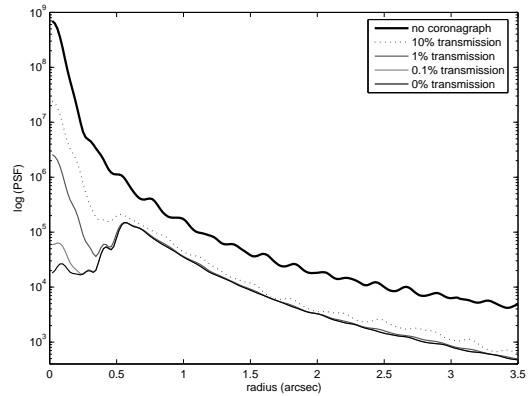,width=8cm,bbllx=31pt,bblly=230pt,bburx=562pt,bbury=610pt}
\caption{Change in coronagraph suppression performance for disc masks 1 arcsec in
diameter with different transmission.}
\label{trans_mask}
\end{center}
\end{figure}

The chromium disc focal plane masks in OSCA were measured to transmit more light at longer
wavelengths.  To model the effect this might have on the suppression performance,
a set of simulations were performed using a 1 arcsecond disc mask function with
transmission values of 0, 0.1, 1 and 10\%.  These transmission levels are equivalent
to neutral density (ND) values of $\infty$, 3, 2 and 1.  A 2000 phase-screen AO simulation with
the coronagraph using an 80,20 (80\% primary masking, 120\% secondary masking)
Lyot stop was run for each of these disc transmissions and the results plotted in
Fig. \ref{trans_mask}.

The suppression factors (no coronagraph/with coronagraph) for these masks
measured at a radius of 1 arcsec (i.e. 0.5 arcsec from the edge of the mask),
starting with the 0\% transmission are: 4.80, 4.80, 4.65 and 3.83 respectively.
Therefore for small amounts of transmission, 1\% and less, there is negligible loss of
suppression performance compared to a totally opaque mask.  For a 10\%
transmission mask the difference becomes significant, with the factor loss in
suppression measured at 1 arcsecond radius being 1.25 compared to the opaque
mask.

\begin{figure}
\begin{center}
\epsfig{file=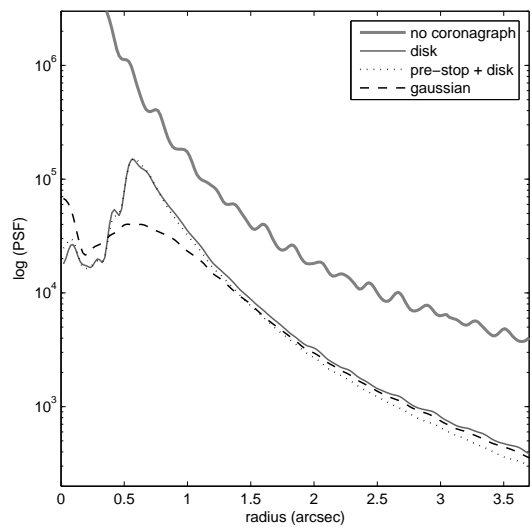,width=8cm,bbllx=97pt,bblly=227pt,bburx=497pt,bbury=614pt}
\caption{Simulated PSF profiles for the different shaped coronagraphic focal masks, the
standard (80,20) Lyot stop was used for each.  Characteristic width for each mask
is 1.0 arcsec.  Phase-screens corrected with a NAOMI-like simulated system - 2000,
$256\times256$ phase-screens with an 4 times FFT padding factor.  Peak for
no-coronagraph lies at $7\times10^8$.}
\label{shaped_masks}
\end{center}
\end{figure}

Simulations were also performed for a wide variety of different shaped focal
masks, Fig. \ref{shaped_masks} shows the simulated suppression profiles of those mask types
most relevant for OSCA.  As expected, the best suppression close in to the mask
(0.5-1.0 arcsec) is achieved by using a Gaussian shaped focal mask.  For
distances $>2$ arcsec the disc mask with a pre-AO stop gives the best
suppression.  The segmented adaptive mirror contributes a significant amount to
the diffracted/scattered light in the image, so reducing the unwanted light
falling on the mirror acts to reduce the background.  The effect of smoothing
the edges of sharp-edged masks and using Gaussian profiled masks improves the
suppression by reducing additional diffracted light in the pupil plane (see the
result of diffraction in the pupil plane by a hard-edged mask in Fig. \ref{pupil_phase}(b)).
This then increases the efficiency of the Lyot mask in removing the light
in the remaining wings of the PSF.  The light from remainder of the PSF,
when viewed at the next pupil plane is then
mostly concentrated about the edges of this aperture image (primary, secondary, vanes,
segment edges etc.), i.e. the parts of the aperture image that contain the most high spatial
frequencies.

\subsection{Effect of a segmented mirror on performance}
\begin{figure}
\begin{center}
\epsfig{file=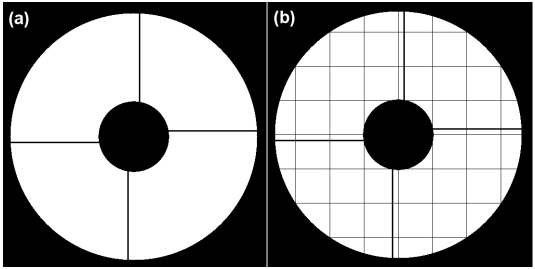,width=7cm,bbllx=20pt,bblly=20pt,bburx=416pt,bbury=218pt}
\caption{(a) Simulated WHT aperture function; (b) as in (a) but with an
approximate NAOMI segmented mirror pattern applied.}
\label{aperture}
\end{center}
\end{figure}

There are a number of additional factors which need to be taken into
consideration when designing a coronagraph with an adaptive segmented mirror.
The segmentation pattern will cause additional unwanted diffraction effects
which distribute light away from the central maximum, this includes the gap
between the segments and any phase effects due to step mismatching between the
segments.  The pupil in OSCA (no focal or Lyot mask) is illustrated in Fig. \ref{aperture}.
The normalised peak intensity ratio of the PSF produced by this aperture
compared to the same aperture
without segmentation (with no phase mismatching) is 0.97.  From this simple
calculation it is seen that a segmented AO system is of lower contrast by design
compared to a continuous face-sheet mirror.

However, the loss in performance when using a coronagraph with a segmented
mirror is much more than 3\%.  The reason for this is can be seen by examining
the distribution of light in the pupil plane after the application of the focal
stop, an example is given in Fig. \ref{pupil_phase}(a).  For the case of a Gaussian focal
plane mask applied to the PSF obtained by using the aperture in Fig. \ref{aperture}(b), it
is found that 45\% of the total light in the pupil is distributed about the
segment edges.  So if a normal Lyot mask is used here (primary, secondary and
vane masking) a significant proportion of the light from that remaining of the
masked star will stay in the final image, thus reducing the suppression
performance.

\begin{figure}
\begin{center}
\epsfig{file=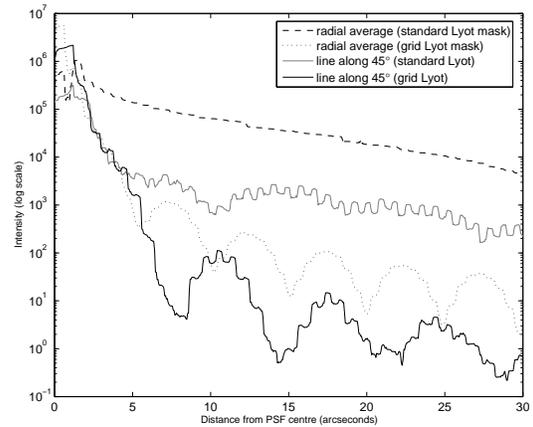,width=8cm,bbllx=0pt,bblly=187pt,bburx=597pt,bbury=653pt}
\caption{Plot showing the benefit of using a grid-type Lyot mask matched to the
NAOMI segment pattern compared to a standard Lyot mask (primary and secondary
masking only) - the intensity has been scaled to the same throughput for both
masks.  Simulations used a 1024 pixel aperture diameter padded to 2048 for the
FFT, 8 segments across the aperture with a 2 pixel gap width.  A 2 arcsec solid
disc was used as the occulting mask.}
\label{grid_plot}
\end{center}
\end{figure}

\begin{figure*}
\begin{center}
\epsfig{file=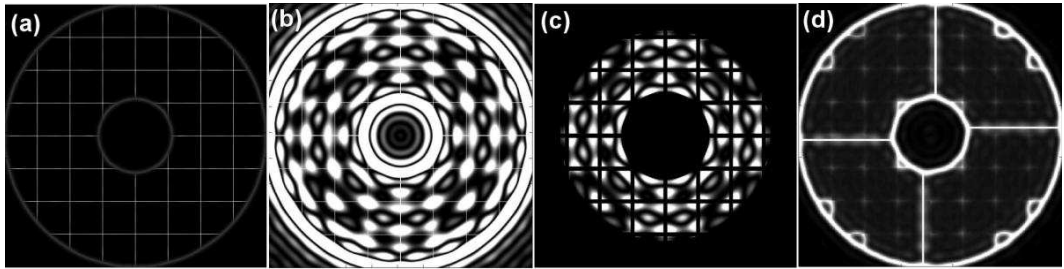,width=14cm,bbllx=20pt,bblly=20pt,bburx=453pt,bbury=129pt}
\caption{Pupil plane images after the application of a 2.0 arcsec hard circular
stop. (a) aperture with segment gap representation of 2 pixels, image has been
scaled to minimum and maximum levels; (b) same image as (a) but cut at a low intensity
level to bring out faint structures; (c) as in (b)
but with the grid-type Lyot mask applied; (d) no gaps, but full AO simulation
showing the effect of phase errors between the segments, the secondary vane
pattern based on the WHT aperture is included.  The outer 4 and inner 4 segments
are turned off in the AO simulation, the increased intensity about these
indicates the greater phase mismatching.  The simulation matches the mid-point
of the segments in piston and so the corners are also at higher intensity due to
phase errors.}
\label{pupil_phase}
\end{center}
\end{figure*}

The use of a more complex Lyot mask which masks the individual mirror segments
as well as the telescope primary and secondary gives improved suppression performance
compared to a Lyot mask which only masks the telescope primary and secondary.
The individual segments in NAOMI are 7.6mm across with a $\sim$0.1mm gap between each
one.  The ratio of gap to segment size is the same order of magnitude to
those proposed for future segmented extremely large telescopes (ELTs), i.e. 1m segments with
10mm gaps.  Hence these results have relevance for high contrast imaging with
ELTs.
To model the effect of the gaps between the NAOMI segments more pixels
are required across the aperture.  The results shown in Fig. \ref{grid_plot}
were obtained using 1024 pixels across the aperture diameter with a 2 pixel gap
between the mirror segments.  Due to the size of this array this was a static
simulation (i.e. not an AO simulation), a 2 times padding factor was used in
the FFTs.  The lines show the mean trends (a convolution filter has been applied
to flatten out the high frequency periodicity) that the segmented aperture
produces.  In the high contrast direction ($45^\circ$ to image axes) the benefit is
most evident at distance $>5$ arcsec from the centre, reducing counts by 2 orders
of magnitude.  For the radial averaged lines (which include the bright axial
diffraction peaks) the benefit of the grid Lyot mask is noticeable from 1 arcsec.

For the full AO simulations (with 256 pixels across the aperture, $4\times$
padding and no gaps),
the effect of phase mismatching between the segments can be seen as an over-intensity
about the segments in the pupil plane subsequent to the application of
the occulting mask, as shown in Fig. \ref{pupil_phase}(d).  Segments with the greatest
intensity are those that are `turned off' and so have the greatest phase step
between them and adjacent segments.  Taking gaps and phase errors between segments
into consideration the benefit of a Lyot mask which masks the individual mirror
segments becomes apparent.

A Lyot mask matched to the NAOMI mirror segments (80\% under sizing of segments)
was created with OSCA (shown in Fig. \ref{lyot} (inset top, left-hand mask)) but is as yet untested on-sky.  The mask
requires very careful alignment and there has been insufficient commissioning
time to trial this new mask.  Telescope schedules allowing, trials may be
performed towards the end of 2005.

\section{Laboratory testing the focal plane masks}

A spatially filtered collimated 613nm laser beam and a series of lenses and masks were
arranged in the laboratory to simulate an ideal coronagraphic system.  A simple iris was used
for the entrance aperture and another one at 80\% diameter to act as the Lyot
mask.  Images were recorded at the final focus for a variety of different occulting
spots using a Santa Barbara Instrument Group (SBIG) camera, this consists of a $375\times242$
pixel, Peltier-cooled CCD.  The usual calibrations were taken - dark frames and
background images between every image.

\begin{figure}
\begin{center}
\epsfig{file=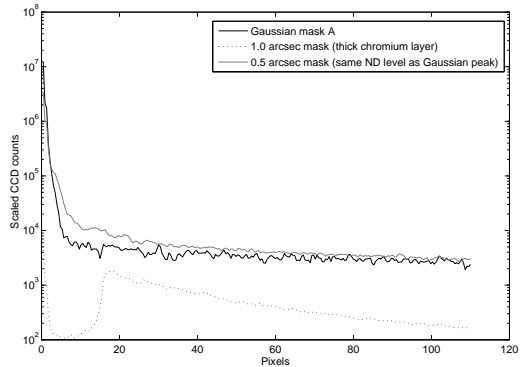,width=8cm,bbllx=-10pt,bblly=214pt,bburx=607pt,bbury=626pt}
\caption{Results of lab testing of focal masks in an ideal coronagraphic system.}
\label{lab_masks}
\end{center}
\end{figure}

These tests were performed after OSCA
had been shipped to the WHT.  The Gaussian masks were commissioned at a later
date and a method to compare them to the standard masks on OSCA was devised.
The lithography template plate was used in place of a 0.5 arcsec disc occulting
mask, approximately the same as the full-width half-maximum of the Gaussian
masks.  The ND value of this mask was also closely matched to the max ND level
at the peak of the Gaussian mask so offered a good comparison between the two
different shapes of mask.  The 1.0 arcsec mask tested here was a spare from OSCA
and had an ND level of $\sim$5.5 (compared to $\sim$2.5 for the 0.5 arcsec mask
and Gaussian) at this wavelength.

Images were taken at the focus for all 3 different occulting masks, both with
and without the Lyot mask.  The images were reduced (dark and background 
subtracted and scaled for exposure differences) and then
radial averages were plotted about the PSF peak.  Fig. \ref{lab_masks} shows
the radial averages of the 3 different masks using the same Lyot mask.  The 1.0
arcsec mask performs best of all, entirely due to its much larger size (covers
$4\times$ the area thus removing much more of the PSF) and greater opacity.
Comparing the other two masks which differ mainly in their shape rather than any
other factors it can be seen that as the simulations predicted, the Gaussian
shaped mask provides greater suppression closer in to the centre than the disc
mask does.

\begin{figure}
\begin{center}
\epsfig{file=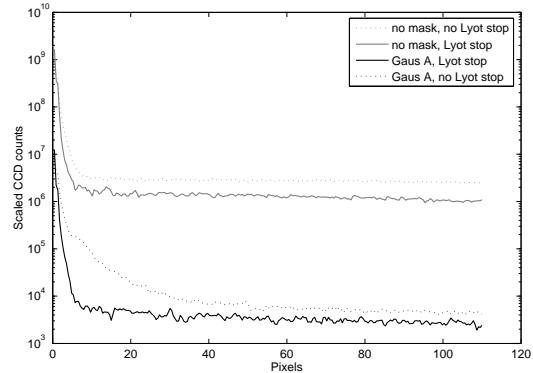,width=8cm,bbllx=19pt,bblly=233pt,bburx=574pt,bbury=607pt}
\caption{Radially averaged line profiles taken about the peak count, for
different coronagraph configurations with the Gaussian focal mask.}
\label{gausa_graph}
\end{center}
\end{figure}

\begin{figure}
\begin{center}
\epsfig{file=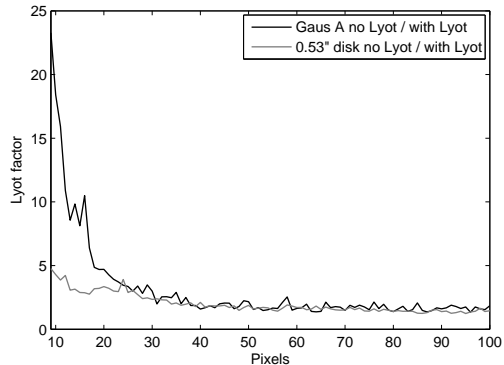,width=7.5cm,bbllx=85pt,bblly=274pt,bburx=509pt,bbury=567pt}
\caption{Plot of `Lyot factors' for the Gaussian and 0.50 arcsec mask, where
Lyot factor denotes ratio between the suppression curves with and without a Lyot
stop.  The x-axis starts just outside the edge of the 0.50 arcsec disc mask.}
\label{lyot_factor}
\end{center}
\end{figure}

Fig. \ref{gausa_graph} shows the significant effect adding a Lyot stop has on
the suppression with the Gaussian mask, the CCD count at 10 pixels from the
centre is $5\times10^3$, compared to $1.2\times10^4$ for the 0.50 arcsec disc.
Without a Lyot stop the value at 10 pixels from the centre for the Gaussian mask
is $8\times10^4$ and for the 0.50 arcsec disc is $5.5\times10^4$.  This shows
that for a Gaussian mask at the focus compared to a solid disc of
comparable size, an undersized pupil stop further along the beam works to
greater advantage in suppressing the final image.  In terms of attenuation
factors, the effect of adding the Lyot stop (measured at 10 pixels) with the
Gaussian mask compared to no Lyot stop brings about a factor of 16 drop in
measured CCD counts, for the 0.50 arcsec disc this factor is only 4.6. At 60
pixels, for the Gaussian the factor is 1.83 and for the 0.50 arcsec disc 1.75,
approximately the same.  The extra effectiveness of the Lyot stop with the
Gaussian mask is therefore greatest close in to the mask, with the `Lyot factor'
dropping as the distance from the mask increases until it converges with the
0.50 arcsec mask values. This trend is shown in Fig. \ref{lyot_factor}.

\section{Commissioning results}

OSCA and its electronics were shipped to La Palma at the
beginning of May 2002 and it underwent first commissioning soon after.
Only two nights on-sky were allocated for the OSCA commissioning so time was
very limited.
The first night was plagued by extremely bad seeing throughout (5
arcsec recorded at worst) and the AO system could not be used.  Since
the largest occulting mask in OSCA is 2.0 arcsec it was also not
feasible to do any performance testing.

\begin{figure}
\begin{center}
\epsfig{file=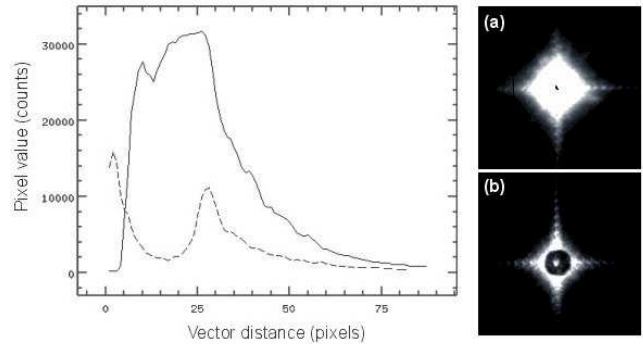,width=8.5cm,bbllx=20pt,bblly=20pt,bburx=446pt,bbury=262pt}
\caption{Cross-sections taken in the high-contrast ($45^\circ$) direction from {\it H}-band images
(a) without OSCA (solid line) and (b) with 2.0 arcsec mask (dashed line).  The
peak at zero is due to the OSCA mask having a small throughput.  The jagged drop
to zero ($< 25$ pixels) with no coronagraph is due to detector saturation.  The INGRID pixel
scale is 0.04 arcsec/pixel.}
\label{may_sup}
\end{center}
\end{figure}

The last night (24th May) saw very variable seeing over the course of
the night and sky-location and the presence of high cirrus cloud also
caused problems on occasion.  NAOMI was used although centring
objects on the OSCA occulting masks was difficult when the seeing was
bad and OSCA performance was degraded.  The average seeing was 1.5
arcsec, and with the AO system an average corrected PSF width of 0.5
arcsec was obtained.  Fig. \ref{may_sup} shows the suppression obtained using
OSCA in {\it H}-band using the 2.0 arcsec mask during these conditions.  From the
graph it can be seen that just outside the edge of the occulting mask the photon 
count has been reduced by a factor of 3.5.  These values have already been 
adjusted to account for the loss in throughput due to the Lyot mask.

Attempts were made to observe science targets during the course
of the night.  The objects were
chosen based on their need for coronagraphic observations, i.e. features that
would not otherwise be easily observable and would demonstrate the benefits of
using a coronagraph.  They also had to have a {\it V}-band magnitude of less than 12
(i.e. brighter) due to the sensitivity of the NAOMI wavefront sensor and be
observable during the night at an altitude greater than $z=40^\circ$; below
this the turbulence is generally higher due to the high air-mass. Additionally two
possible subtraction stars were found for each target.  These were selected to
be as close to the target in all respects -- sky position, colour/spectral type
and {\it V} magnitude.  The attempt was to try and select single stars (no known
companions) with a 10 arcsec field about them that is free of any other
(particularly bright) stars.  Finding good subtraction stars is difficult as
there are no comprehensive catalogues for this and information can be
incomplete or incorrect in the ones that are available, the ones discussed here were
chosen with the aid of the SIMBAD database.

\begin{figure}
\begin{center}
\epsfig{file=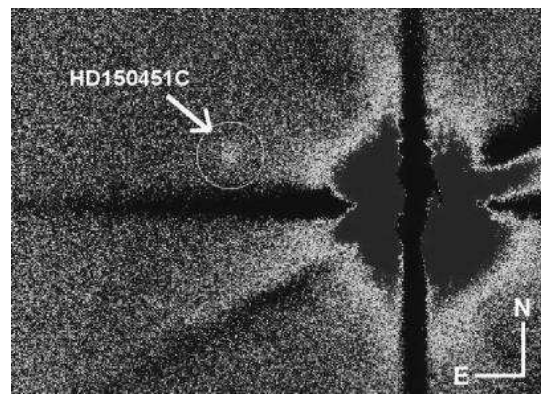,width=7cm,bbllx=20pt,bblly=20pt,bburx=164pt,bbury=126pt}
\caption{Detection of HD150451C (white dwarf) using NAOMI+OSCA.  The central star
has had a scaled calibration star PSF subtracted from it to increase contrast.}
\label{obj15}
\end{center}
\end{figure}

The reduced data for one of the commissioning targets -- HD150451AB -- revealed a faint
detection (see Fig. \ref{obj15}) of the
recently identified cool white dwarf companion HD150451C \citep{car05}.  At the time a
potential brown dwarf companion was suspected.
Since no field rotations were performed to confirm this was not an
AO artefact and the signal to noise for the dwarf was extremely low, no specific
conclusions could be made as of May 2002.  The data collected over 2 years by \cite{car05}
has confirmed this to be a companion to HD150451AB, and although initial data
suggested the companion to be a methane brown dwarf, recent spectroscopic
measurements favour a cool white dwarf classification.

\begin{figure}
\begin{center}
\epsfig{file=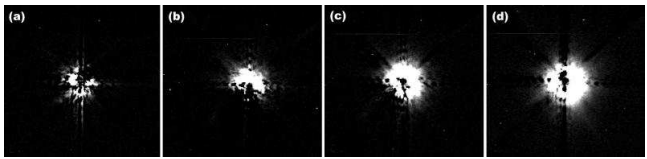,width=8.5cm,bbllx=20pt,bblly=20pt,bburx=308pt,bbury=88pt}
\caption{A series of same object subtractions over time in {\it J}-band to illustrate
the changing PSF due to NAOMI.  (a) after 1 minute, (b) 5 minutes, (c) 15
minutes and (d) 40 minutes.  Atmospheric seeing conditions were average (0.7
arcsec in {\it V}) and each of the images used in the subtraction had an integration
time of 30 seconds.  All images are scaled to same level.}
\label{residuals}
\end{center}
\end{figure}

Sky frames were taken for all data (every 10 minutes in {\it K} and every 15 minutes
in {\it J}) to allow more accurate background subtractions.  Two different PSF subtraction
stars were observed for each target, although with much longer intervals than is
ideal due to time constraints.  It became apparent that the PSF stability was poor
so that subtractions across timescales greater than 5 minutes were
contaminated with many AO residuals.  Fig. \ref{residuals} shows this PSF
change over time, taken from {\it J}-band images of HD141569.

\begin{figure}
\begin{center}
\epsfig{file=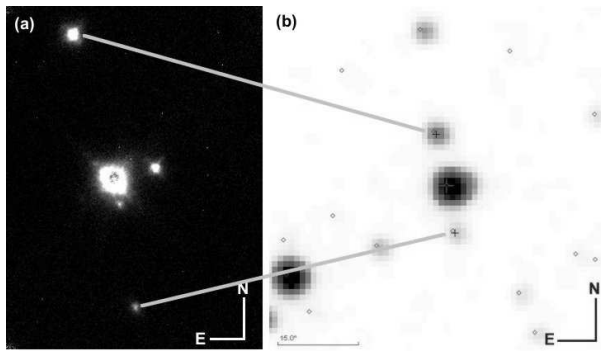,width=8cm,bbllx=20pt,bblly=20pt,bburx=231pt,bbury=141pt}
\caption{$K_s$-band image of BD-04 4476 (SAO 142339).  (a) There are 4 stars
about the central (occulted) star at distances of 1.59, 2.72, 8.26 and 9.67
arcseconds.  (b) 2MASS image of the same region.  The two stars marked with
crosses are the outermost stars in the OSCA image.}
\label{psf5}
\end{center}
\end{figure}

Due to the limited time, nothing of interest was uncovered in data taken on MWC297.
However the PSF star chosen for MWC297 revealed a group of 4 stars in close
proximity (see Fig. \ref{psf5}(a)), with the closest being 1.6 arcseconds
from the central PSF star.  No off-mask images were taken for this object since
it was originally only to be used as a PSF subtraction star, so only rough estimates of
the {\it K}-magnitude were possible.  The occulted profile was compared to that of an
un-masked IR standard star and scaled so that the wings closely matched.  This
scale factor (corrected for integration time) was then used directly with the IR
standard star magnitude and applied to the usual magnitude-flux relation to
obtain a value of 8.9.  The {\it K}-magnitude estimate for the closest star is $12.2 \pm 0.3$ --
a value for the flux contribution from the central star has been subtracted and
was estimated by plotting radial averages about the star.  The central star and
the outermost two stars in this image are listed in the latest Two Micron All Sky Survey
(2MASS) catalogue of
point sources.  This lists BD-04 4476 as having a {\it K} magnitude of 8.5 and the
outer two stars 12.6 and 11.3 respectively.  Taking into account the estimates
involved with the focal stop and that 2MASS cannot
resolve the inner two stars, our measurements appear to be reasonable.

An upgrade to OSCA was carried out in April 2003.  The Gaussian occulting masks
were installed along with a razor edged anti-scatter mask in front of the focal
plane substrates (to reduce scattering from the edges of substrates on which the
occulting spots are deposited).  The whole of OSCA was also moved and installed at
GRACE (a new, environment controlled laboratory on one of the Nasmyth platforms
on the WHT).  This should alleviate previous problems of dust contamination and 
temperature problems for NAOMI.

\section{Conclusions}

OSCA is a high precision
stellar coronagraph, produced on a low budget, over a short timescale and meeting all the
design specifications. However, due to the limited time assigned for the 
commissioning and the overlap with NAOMI engineering schedules, a complete and 
thorough testing of OSCA has not been possible and as a result there have been 
no performance tests done with OSCA in optimum seeing conditions {\it and} optimum 
alignment (both of OSCA and other instrumentation).

The mechanics and electronics of OSCA have operated consistently to date and succeed in
maintaining the required positioning and alignment, including that for the Lyot
stop rotation.
The deployment mechanism for OSCA has proven to be very successful; it allows
OSCA to be raised in to (and out of) the beam very quickly and with consistently
accurate positioning, allowing for a flexible and varied observing program over
the course of a night.    

Simulations of the NAOMI and OSCA system also have relevance for the next
generation of extremely large telescopes where high contrast imaging is 
essential in the search for extra-solar planets.  When designing a coronagraphic
system for a highly segmented aperture suitable masking in the Lyot (pupil)
plane must be devised to counteract the strong diffraction pattern that will
result otherwise - reducing suppression performance and confusing the data.

A number of interesting astronomy targets have been observed during the OSCA
commissioning runs.  The most positive results were the very faint detection of the
cool white dwarf in HD150451 and the discovery of a potential companion
($\sim 150$AU) in BD-04 4476.  More data is required to draw any firm
conclusions on the nature of these objects.

Information regarding the current status of OSCA and instructions for observing 
with the system can be found on the ING website.

\section*{Acknowledgments}

In memory of Richard Bingham, who passed away during the publication of this paper.
Richard was a great colleague and a brilliant optical designer and will be sadly missed
by us all.

We would like to thank the Isaac Newton Group at the William Herschel Telescope
for their support during the commissioning of OSCA.  We also thank the
astronomers at University College London and elsewhere that gave us help and
suggestions for observing and objects of scientific interest.  The funding for
OSCA was provided by the UK Particle Physics and Astronomy Research Council
(PPARC).  SJT acknowledges PPARC for providing the Ph.D studentship, during
which the research for this paper was completed.

This research has made use of the SIMBAD database, operated at CDS, Strasbourg, France

\bsp

\label{lastpage}

\end{document}